\documentclass[a4paper, 12pt]{article}
\usepackage{epsfig}
\usepackage{graphicx}
\usepackage{amsmath}

\author{Juli\'an Candia$^{a}$ and Karina I. Mazzitello$^{b}$\\{}\\
$^a${\small\it Center for Complex Network Research and Department of Physics,}\\  
{\small\it Northeastern University, Boston, MA 02115, USA}\\  
{\small Email address: jcandia@nd.edu}\\
$^b${\small\it CONICET and Departamento de F\'{\i}sica, Facultad de Ingenier\'{\i}a,}\\  
{\small\it Universidad Nacional de Mar del Plata, Mar del Plata, Argentina}\\
{\small Email address: kmazzite@mdp.edu.ar}}

\title{Mass Media Influence Spreading in Social Networks with Community Structure} 

\begin{document}
\maketitle

\begin{abstract}
We study an extension of Axelrod's model for social influence, in which  
cultural drift is represented as random perturbations, while mass media are introduced by means of 
an external field. In this scenario, we investigate how the modular structure of social 
networks affects the propagation of mass media messages across the society.  
The community structure of social networks is represented by coupled random networks, 
in which two random graphs are connected by intercommunity links.  
Considering inhomogeneous mass media fields, we study the conditions for 
successful message spreading and find a novel phase diagram in the multidimensional parameter space.   
These findings show that social modularity effects are of paramount importance in order to design 
successful, cost-effective advertising campaigns.
\end{abstract}

\section{Introduction}
Over the last few years, statistical physics has increasingly contributed useful tools and valuable insight into 
many emerging interdisciplinary fields of science \cite{oli99,wei00,sta06}. 
In particular, many efforts have focused recently on the mathematical modeling 
of a rich variety of social phenomena, such as social influence and self-organization, cooperation, 
opinion formation and spreading, evolution of social structures, etc 
(see e.g. \cite{gal00,szn00,ber01,ber02,kup02,ale02,sch03,ben05,smi06,des06,can06,can07a,ang07,bor07a,bor07b,can07b,cas07}).

In this context, an agent-based model for social influence, originally proposed by Axelrod \cite{axe97a,axe97b} 
to address the formation of cultural domains, has been extensively studied within the sociophysics community 
(see e.g. \cite{cas00, kle03a, kle03b, gon05, gon06, kup06, maz07, gon07}). 
In Axelrod's model, culture is defined by the set of cultural attributes (such as language, art, 
technical standards, and social norms \cite{axe97b}) subject to social influence. The cultural state of 
an individual is given by their set of specific traits, which are  
capable of changing due to interactions with their acquaintances. In the original formulation, the individuals are 
located at the nodes of a regular lattice and the interactions are assumed to take place between lattice neighbors.  
Social influence is defined by a simple local dynamics, which is assumed to satisfy the following two properties: 
(a) social interaction more likely takes place between individuals that share some of their cultural traits; 
(b) as a result of the interaction, their cultural similarity is increased. 

Earlier investigations showed that the model undergoes a phase transition separating an 
ordered (culturally polarized) phase from a disordered (culturally fragmented) one, 
which was found to depend on the number of different cultural traits available \cite{cas00}.  
The critical behavior of the model was also studied in different complex network topologies, such as small-world and 
scale-free networks \cite{kle03a}. 
These studies considered, however, zero-temperature dynamics that 
neglected the effect of fluctuations.
Following Axelrod's original idea of incorporating random perturbations to describe the effect 
of {\it cultural drift} \cite{axe97a}, noise was 
later added to the dynamics of the system \cite{kle03b}. 
With the inclusion of this new ingredient, the disordered multicultural configurations were found to be metastable 
states that could eventually decay towards ordered stable configurations, depending on the competition between the noise rate 
and the characteristic time for the relaxation of perturbations.

Very recently, other extensions of the model were proposed, in which the role of mass media was investigated within 
different scenarios. Neglecting random fluctuations, some studies considered in detail the role of 
external \cite{gon05} and autonomous local or global fields \cite{gon06}. 
Another recent investigation focused on the interplay and competition between 
cultural drift and mass media effects \cite{maz07}. Adopting a mass media coupling capable of 
affecting the cultural traits of any individual in the society (including those who do not share any 
features with the external message), it was shown that the external field can induce cultural ordering 
and reproduce the trend of actual advertising campaign data. 

In a related context, recent investigations addressed the role played by the underlying topology of 
complex substrates on the dynamical and critical behavior of the models defined on them. 
The effects of some structural properties 
that characterize disordered substrates, such as the small-world effect, 
the degree distribution, the degree-degree correlations, and the local clustering, were extensively 
studied \cite{wat99,alb02,new02,dor03,new06}.
 
Furthermore, the property of community structure, or large-scale clustering, appears to be common to 
many real-world networks and is nowadays being subject of intense research efforts. In many social networks, as e.g. 
the well known karate club study of Zachary \cite{gir02}, the United States 
House of Representatives \cite{por06}, scientific co-authorships and mobile phone call records \cite{pal07},  
well defined modular structures were observed.  
However, the effects of community structure on models of sociophysical interest have received so far little attention. 
Lambiotte and Ausloos \cite{lam07} have very recently considered the effect of communities on the Majority Rule model
by means of the so-called {\it coupled random networks}, a mixture of two random communities, in which a  
parameter $\nu$ controls the degree of intercommunity links relative to that of intracommunity connections \cite{gir02}.
Depending on $\nu$ and on a noise parameter, a diagram with three distinct phases is obtained: 
a disordered phase, where no collective phenomena takes place; an ordered, symmetric phase, where both communities 
share the same average state; and an ordered, asymmetric phase, in which different communities reach different states. 
  
The aim of this work is to investigate effects arising from the characteristic modular structure of social networks
in the propagation of mass media messages across the society. To this end, we focus on the extension of Axelrod's 
model proposed in Ref.~\cite{maz07}, which includes effects of mass media and cultural drift, using  
coupled random networks for the substrate. In the absence of external messages, a phase diagram with three phases is found, 
qualitatively analogous to that observed in Ref.~\cite{lam07} for the majority rule model. Then, we assume that an 
inhomogeneous mass media field affects one of the communities and study the system's response to the spreading of the message.
Incorporating the intensity of the mass media field as an additional parameter, 
several new phases are observed to emerge, thus leading to a very rich, novel phase diagram in the multidimensional space of 
model parameters.  

This paper is organized as follows: 
in Section 2, details on the model and the simulation method are given; 
Section 3 is devoted to the presentation and discussion of the results, while Section 4 contains the conclusions. 

\section{The model and the simulation method}

In order to represent the community structure observed in social networks, we consider a substrate topology 
consisting of two coupled random networks (CRN). These structures were first proposed in Ref.~\cite{gir02} 
to carry out comparative tests of different methods for community detection in complex networks. 
We assume that a system of $N$ nodes is divided into two communities ($A$ and $B$) of equal size. A CRN configuration 
is built by adding intracommunity links between pairs of nodes that belong to the same community, as well as intercommunity 
links between pairs of nodes that belong to different communities. 
Considering all possible node pairs, intracommunity links are added with probability $p_{int}$, while 
intercommunity connections are added with probability $p_{ext}$. 
On average, a node is thus connected to $k_{int}= p_{int}(N/2-1)$ neighbors inside the same community and to 
$k_{ext}= p_{ext}N/2$ nodes that belong to a different community. For the sake of simplicity, we fix $k_{int}=4$ 
and tune the intercommunity connectedness by means of a single parameter, namely $\nu\equiv 
k_{ext}/k_{int}\approx p_{ext}/p_{int}$. Notice that the $\nu\to 1$ limit corresponds to a single random graph 
lacking modular features, while $\nu\ll 1$ is the case in which well defined communities are sparsely 
connected with each other. 

The nodes of the system are labeled with an index $i$ ($1\leq i\leq N$) and represent individuals subject to interactions 
with their neighbors (i.e. other individuals directly linked to him/her by either intra- or inter-community connections), 
as well as with an externally broadcast mass media message. According to Axelrod's model, 
the cultural state of the $i-$th individual is described by the integer vector 
${\bf \sigma_i} = (\sigma_{i1},\sigma_{i2},...,\sigma_{iF})$, where 
$1\leq\sigma_{if}\leq q$. The dimension of this vector, $F$, defines the 
number of cultural attributes, while $q$ corresponds to the number of different cultural traits per attribute. 
Initially, the specific traits for each individual are assigned randomly with a uniform distribution. Similarly,  
the mass media cultural message is modeled by a constant integer vector 
${\bf \mu} = (\mu_1,\mu_2,...,\mu_F)$, which can be chosen as ${\bf \mu} = (1,1,...,1)$ without loss of generality. 
The intensity of the mass media message relative to the local interactions between neighboring 
individuals is controlled by the parameter $M$ ($0\leq M\leq 1$). Moreover, the parameter $r$ ($0 < r\leq 1$) 
is introduced to represent the noise rate \cite{kle03a}. 

Since the main focus of this work is on mass media spreading phenomena under the influence 
of an underlying modular substrate, we will consider inhomogeneous mass media affecting only 
individuals that belong to the community $A$. 
The model dynamics is defined by iterating a sequence of rules, as follows: (1) an individual
is selected at random; (2a) if the individual belongs to the community $A$, he/she interacts with the mass media 
field with probability $M$, while he/she interacts with a randomly chosen neighbor with probability (1-$M$); 
(2b) if the individual belongs to the community $B$, he/she interacts with a randomly chosen neighbor;
(3) with probability $r$, a random single-feature perturbation is performed.  

The interaction between the $i-$th and $j-$th individuals is governed by their cultural overlap,  
$C_{ij}=\sum_{f=1}^F\delta_{\sigma_{if},\sigma_{jf}}/F$, where $\delta_{kl}$ is the Kronecker delta. 
With probability $C_{ij}$, the result of the interaction is that of 
increasing their similarity: one chooses at random one of the attributes on which they differ 
(i.e., such that $\sigma_{if}\neq\sigma_{jf}$) and sets them equal by changing one of their traits.
Naturally, if $C_{ij}=1$, 
the cultural states of both individuals are already identical, and the interaction leaves them unchanged. 

The interaction between the $i-$th individual and the mass media field is governed by the overlap term 
$C_{iM}=(\sum_{f=1}^F\delta_{\sigma_{if},\mu_f}+1)/(F+1)$. Analogously to the precedent case, 
$C_{iM}$ is the probability that, as a result of the interaction, the individual changes one of the traits 
that differ from the message by setting it equal to the message's trait. 
Again, if $C_{iM}=1$, the cultural state of the individual is already identical to the mass media message, 
and the interaction leaves it unchanged. 
Notice that $C_{iM}>0$; thus, the mass media coupling used here is 
capable of affecting the cultural traits of any individual within community $A$, 
including those who do not share any features with the external message. 

As regards the perturbations introduced in step (3),
a single feature of a single individual is randomly chosen, and, with probability $r$, 
their corresponding trait is changed to a randomly selected value between 1 and $q$. 

In the absence of fluctuations, the system evolves towards absorbing states, i.e., frozen configurations that 
are not capable of further changes. However, for $r>0$ the system evolves continuously and, after a transient period, 
it attains a stationary state. Following previous studies on Axelrod's model \cite{gon05,maz07}, 
in this work we chiefly focus on systems of fixed size ($N=2500$ nodes), fixed number of cultural attributes ($F=10$) 
and fixed number of different cultural traits per attribute ($q=40$). Furthermore, we also briefly discuss the effects 
(or lack thereof) observed by changing these model parameters.   
The results presented in the next Section correspond to observables measured over statistically-averaged ensembles in 
the stationary regime, which were obtained by averaging over 200 different (randomly generated) initial 
configurations and 100 different network realizations.  

\section{Results and discussion} 
In order to set the stage for the investigation of modularity effects,  
let us first briefly summarize the main results concerning Axelrod's model defined on the square lattice.
As mentioned above, in the absence of fluctuations the system reaches absorbing configurations, in 
which the state of each individual is fixed and not capable of further changes.
With the inclusion of noise to model the effect of cultural drift, however, 
disordered multicultural configurations become metastable 
states that can eventually decay to ordered stable configurations \cite{kle03b}.
Whether this decay actually takes place or not depends   
on the competition between the noise rate, $r$, and the 
characteristic time for the relaxation of perturbations, $T$. For $r<T^{-1}$, repeated cycles of perturbation-relaxation 
processes drive the disordered system towards monocultural states, while, for $r>T^{-1}$, noise 
rates are large enough to hinder the relaxation mechanism, thus conserving the disorder. With arguments based on 
a mean-field description of a damage spreading process, the characteristic time for the relaxation of perturbations 
is estimated as $T\sim N\ {\rm ln}N$, where $N$ is the system size \cite{kle03b}. 

In the absence of noise, the number of cultural traits is observed to play a key role in determining the final 
absorbing state: ordered monocultural configurations (for $q<q_c$) and disordered multicultural 
ones (for $q>q_c$) are separated by 
a finite critical value $q_c>0$ \cite{cas00,kle03b}. For instance, in a system of size $N=2500$ with 
$F=10$ cultural attributes, the transition takes place at $q_c\approx 50$. 
For $r>0$, however, the order-disorder transition solely depends on
the effective noise rate $r_{eff}=r\times (1-1/q)$. The very mild dependence on the parameter $q$ 
just stems from the fact that, according to the third rule of the model dynamics, 
a perturbation can leave the cultural configuration unchanged with probability $1/q$. For the $N=2500$ and 
$F=10$ case, the order-disorder transition is observed around $r_{eff}\approx 10^{-4}$ \cite{kle03b}. 
  
When both noise and the mass media external field, $M$, are taken into account, interplay and competition effects 
are observed \cite{maz07}. As the field intensity is increased, the transition shifts to higher noise levels.   
Since the ordering is driven by the mass media field, the system attains a unique 
ordered state, namely, the monocultural state in which all individuals share their 
cultural traits with those of ${\bf \mu}$, the external message. In the absence of external fields, 
the noise-induced ordering leads to $q^F$ equally likely monocultural 
ground configurations, as well as to excursions from a ground configuration to another one. 
Note hence that, due to considerations of ergodicity and 
the multiplicity of ground states, it is not possible to simply define an 
``effective noise intensity" $r^\prime=r^\prime(r,M)$ in order to trivially map 
the model with field onto an effective model without field.

Let us now focus on the effects of modularity, which is modeled using a substrate topology that
consists of two coupled random networks (see Sect. 2 for details). 
Firstly, we will address cultural drift effects alone ($r>0, M=0$); later, we will study the case in which 
inhomogeneous mass media affect one of the communities ($r>0, M>0$) and explore the conditions for successful message 
spreading across the whole system.  

\begin{figure}[t!]
\begin{center}
\epsfxsize=4.2truein\epsfysize=2.9truein\epsffile{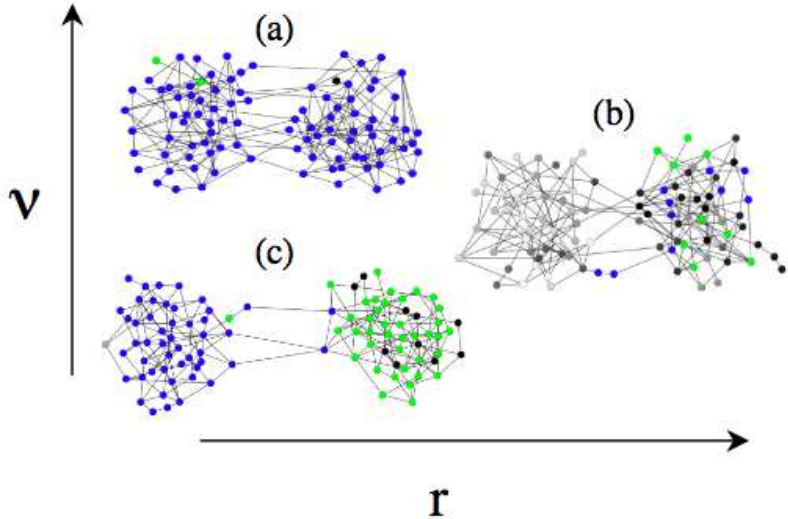}
\end{center}
\caption{Typical snapshot configurations for different values of noise, $r$, 
and intercommunity connectedness, $\nu$, in the absence of mass media fields ($M=0$). 
Nodes belonging to the community $A$ ($B$) are shown on the left (right) side of each CRN realization. 
The most popular cultural state is shown in blue, the second most popular in green, 
the rest in greyscale. The network visualizations were created with Cytoscape \cite{sha03}.}
\label{fig1}
\end{figure}
 
Figure 1 presents some typical snapshot configurations of the stationary regime for 
different values of noise, $r$, and intercommunity connectedness, $\nu$, in the absence of mass media fields: 
(a) $r=10^{-3}, \nu=6\times 10^{-2}$; (b) $r=10^{-2}, \nu=4\times 10^{-2}$; and (c) $r=10^{-3}, \nu=1.5\times 10^{-2}$. 
Here and throughout, the community $A$ ($B$) is shown on the left (right) side of each network realization. 
For the sake of clarity, snapshot visualizations correspond to networks of 
small size ($N=100$). The cultural state of the individuals is indicated by different node colors: the cultural 
state shared by the largest number of individuals is shown in blue, the second most popular cultural state is shown 
in green, while less frequent states are indicated in greyscale.
    
The characteristic configuration for small noise levels and many intercommunity links (Figure 1(a)) is a nearly full 
consensus: most of the individuals share the same cultural state. However, when considering larger noise 
rates (Figure 1(b)), the system undergoes a transition towards complete disorder, where the size of the most popular 
cultural state represents just a small fraction of the total system's size. Indeed, these phenomena are reminiscent of the 
observed behavior of Axelrod's model in the lattice, where a finite critical noise, $r_c$, was found to separate 
the ordered, monocultural phase (for $r<r_c$) from the disordered, multicultural one (for $r>r_c$) \cite{kle03b}. 
However, strong effects arising from the modular 
structure of the substrate are observed at small values of $\nu$, leading to the appearance of a new phase 
(Figure 1(c)). This is the ordered, bicultural phase, in which different cultural states prevail within each community.      
Interestingly, this behavior is in qualitative agreement with related work on a 2-state, majority rule model defined on 
substrates with community structure, where the coexistence of opposite opinions was observed \cite{lam07,lam07b}.      

In order to quantitatively characterize different phases in the stationary regime, 
we define $A_{max}$ ($B_{max}$) as the maximal number of members of community $A$ ($B$) 
that share the same cultural state, normalized to unity. Furthermore, we define the vector ${\bf{a_{max}}}$ 
(${\bf{b_{max}}}$) as the prevailing 
cultural state within community $A$ ($B$). Once a stationary configuration is generated, we 
classify the cultural state of 
each community as being ordered or disordered according to a simple majority 
criterion: $A$ ($B$) is ordered if $A_{max}\geq 0.5$ ($B_{max}\geq 0.5$), 
and disordered otherwise. Moreover, when both communities are ordered, the whole system is in an ordered {\it symmetric} state 
if ${\bf{a_{max}}}={\bf{b_{max}}}$, while it is in an ordered {\it asymmetric} state otherwise. 
Since $A$ and $B$ are indistinguishable communities, the combination 
of these different states leads to 4 possible phases. 

Figure 2 shows the resulting phase diagram in the $r-\nu$ parameter space, in which the dominant phases for each region 
are displayed. At any given point on the $r-\nu$ plane, the {\it dominant phase} is defined as the phase with the largest 
probability of occurrence (for instance, phase probability profiles for the $M>0$ case are shown in Figures 4 and 7 below). 
Thus, boundaries separating two phases correspond to states for which two dominant phases are equally probable, while triple points 
are associated to states for which three dominant phases are equally probable. The same definition was also adopted to 
determine the phase diagrams presented below (see Figures 5 and 8). 
   
As anticipated, three distinct regimes prevail: a multicultural (disordered) 
phase, a monocultural phase in which both communities share the same cultural state, and a bicultural phase 
in which each community is ordered, but in different states independent of each other. 
The effect of increasing $r$ at a fixed value of $\nu$ is that of increasing the disorder in the system, as expected. 
The noise-induced order-disorder transition is only mildly dependent on the number of links connecting both 
communities, with the transition curve located at $r\simeq 2\times 10^{-4}$ for $\nu\geq 1.4\times 10^{-3}$. 
The mild $\nu$-dependence of the order-disorder transition curve is due to finite-size effects: 
$\nu$ plays here the role of tuning the ``effective system size" from $N_{eff}=N/2$ 
(in the $\nu\to 0$ limit, when the two communities are independent of each other) to $N_{eff}=N$ 
(in the $\nu\to 1$ limit, when the modular structure washes out).

\begin{figure*}[t]
\begin{center}
\epsfxsize=3.8truein\epsfysize=2.7truein\epsffile{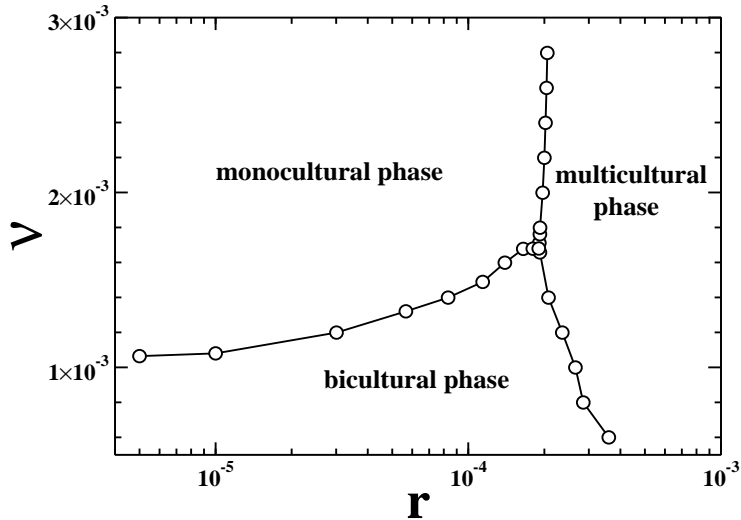}
\end{center}
\caption{Phase diagram for $M=0$ in the $r-\nu$ parameter space.}
\label{fig2}
\end{figure*}

It is within the region with small $r$ where community structure effects become more noticeable.  
For small $\nu$, 
the two communities are weakly connected and do not influence each other, leading to the prevalence of the 
(ordered asymmetric) bicultural phase. However, increasing $\nu$ the noise-driven ordered phases tend towards consensus, 
thus leading to a dominant (ordered symmetric) monocultural phase. A tri-critical point is found at $(r=2\times 10^{-4},
\nu=1.65\times 10^{-3})$.
As commented above, a qualitatively similar phase diagram was obtained in previous investigations of a 2-state  
majority rule model defined on substrates with community structure \cite{lam07,lam07b}. 

The transition from the bicultural phase to the monocultural phase that takes place in the small-$r$ region can be 
roughly estimated by the following theoretical argument. Under conditions of small rate of perturbations, we can assume 
that typically most of the nodes in the community $A$ will tend to agree in the same cultural state $\sigma_A$ (randomly 
chosen among any of the $q^F$ possible cultural vectors), and similarly, most of the nodes in the community $B$ will 
share the cultural state $\sigma_B$. The monocultural phase, $\sigma_A=\sigma_B$, is driven by interactions between pairs of 
border nodes, i.e. those with intercommunity links. If a border node is chosen by rule (1) of the model dynamics, 
its interacting neighbor, chosen in turn by rule (2), has a probability $P_{ext}=k_{ext}/(k_{int}+k_{ext})$ to belong 
to a different community. If $L_\nu$ is the total number of intercommunity links, we can assume that the 
interaction between communities is effectively present for $P_{ext}L_\nu>1$. Thus, $P_{ext}L_\nu\sim 1$ can be 
taken as a rough estimate for the occurrence of the monocultural/bicultural phase transition in the small-$r$ region. 
Since, on average, $k_{ext}=\nu k_{int}$ and $\nu\ll 1$, a border node has $k_{ext}=1$, i.e. only one external neighbor. 
Using $L_\nu=2\nu N$, the condition $P_{ext}L_\nu\sim 1$ reads
\begin{equation}
\frac{2\nu N}{k_{int}+1}\sim 1\ .
\label{estimate}
\end{equation}   
Replacing $k_{int}=4$ and $N=2500$, we obtain $\nu\sim 10^{-3}$, which provides a good estimate for the boundary 
observed  in Figure 2 between the monocultural and bicultural phases. 
An immediate consequence of this simple theoretical argument is that results for different network sizes 
should scale with $\nu$ and $N$ through $L_\nu\propto \nu N$. This predicted behavior was indeed confirmed by 
our simulations of systems of different size. Moreover, increasing $N$ also shifts the boundary between ordered 
phases and the multicultural phase towards smaller values of noise, which is consistent with previous observations 
of noise-driven transitions in Axelrod's model defined on the square lattice \cite{kle03b}.       

Let us now address the case in which mass media affect one of the communities
and explore the conditions for successful message spreading across the whole system. 
In order to capture the characteristic behavior of this system in 
the multidimensional parameter space, we consider separately the small-$r$ ordered region 
and the large-$r$ disordered case. 
Since $r$ can be regarded as a measure of the intrinsic individual determination or ``free will" 
relative to the influence 
exerted by neighbors and mass media, the small-$r$ scenario represents a society with individuals subject to 
strong social pressure,  while the large-$r$ case corresponds to a society characterized by loose 
social ties. Within these different scenarios, the 
adoption of inhomogeneous mass media fields that introduce a physical distinction between the dynamics of 
both communities will drive the system across different phase transitions.

\begin{figure}[t]
\begin{center}
\epsfxsize=4.4truein\epsfysize=2.5truein\epsffile{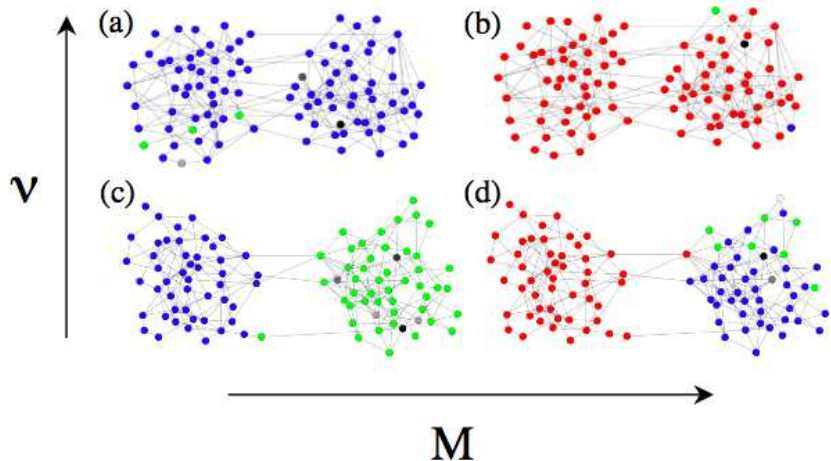}
\caption{Typical snapshot configurations for different values of $M$ and $\nu$ within the small-$r$ regime.
The $\mu$-state is shown in red, the most popular among non-${\bf\mu}$ states appears in blue, 
the second most popular non-${\bf\mu}$ state in green, while other states are shown in greyscale.}
\end{center}
\label{fig3}
\end{figure}

Figure 3 shows typical snapshot configurations of size $N=100$ for different values of $M$ and $\nu$ within the small-$r$ regime 
($r=10^{-3}$): (a) $M=10^{-3}, \nu=6\times 10^{-2}$; (b) $M=10^{-2}, \nu=6\times 10^{-2}$; 
(c) $M=10^{-3}, \nu=2\times 10^{-2}$; and (d) $M=10^{-2}, \nu=2\times 10^{-2}$. 
Recall that nodes belonging to the community $A$ ($B$) are shown on the left (right) side of each network realization. 
The cultural state that 
corresponds to the external message, ${\bf\mu}$, is shown in red. Other states are shown in 
blue (most popular among non-${\bf\mu}$ states), green (second most popular) and greyscale (all other states). 

When communities are strongly interconnected, the system evolves towards consensus, where 
most of the individuals share the same cultural state (Figures 3(a)-(b)). However, the nature of the attained consensus 
depends on the strength of the mass media field: for small $M$, the system organizes itself into any of 
the $q^F$ possible monocultural states, while increasing $M$ a transition takes place towards a regime dominated by 
the mass media message. When communities are instead sparsely interconnected, they tend to evolve independently 
of each other (Figures 3(c)-(d)). However, analogously to the case where communities are tightly bound together, 
the system undergoes an $M$-driven transition from a regime where communities are ordered but independently 
organized into different non-$\mu$ states (Figure 3(c)) to a phase where the mass 
media message prevails within community $A$, while the community $B$ is in a non-$\mu$ state (Figure 3(d)). 
Notice that, lacking enough intercommunity links, even a very intense mass media campaign will fail to convey 
its message to the whole society. 

\begin{figure}[t]
\begin{center}
\epsfxsize=4.8truein\epsfysize=2.5truein\epsffile{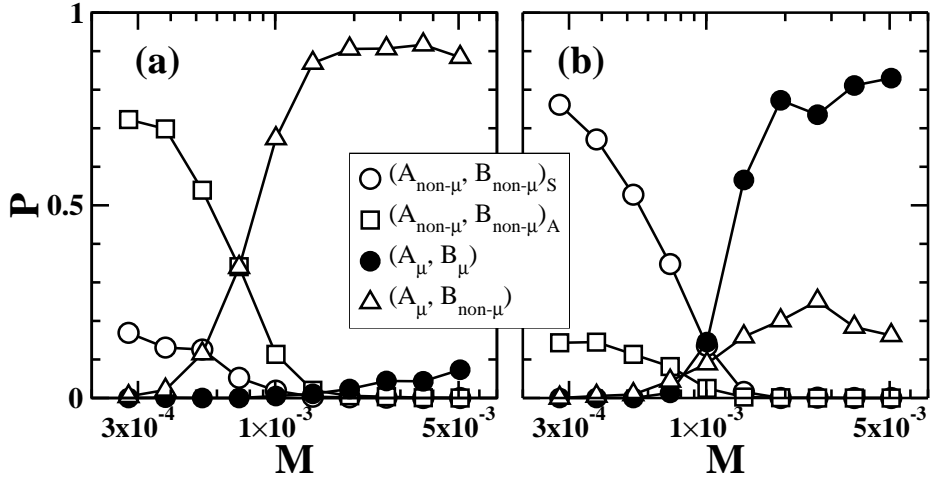}
\caption{Probability of occurrence of the relevant phases as a function of $M$ for  
the small-$r$ regime (with $r=10^{-5}$) and different 
values of the community connectivity: (a) $\nu=6\times 10^{-4}$ and (b) $\nu=2\times 10^{-3}$.}
\label{fig4}
\end{center}
\end{figure}

In order to quantitatively distinguish among different phases, we can follow a procedure similar to 
that described above for the $M=0$ case. 
However, additional phases now arise from the fact that the cultural state corresponding to the mass media message, $\mu$, 
is physically distinguishable from the other $q^F-1$ possible cultural states. For instance, the community $A$ can 
be either dominated by the mass media message ($A_{max}\geq 0.5$ and ${\bf{a_{max}=\mu}}$), ordered in a 
different cultural state ($A_{max}\geq 0.5$ and ${\bf{a_{max}}}\neq\mu$), or disordered ($A_{max} < 0.5$). 
Taking also into account the distinction between symmetric and asymmetric ordered states (which is relevant 
when $A$ and $B$ are ordered in cultural states both different from $\mu$), this ultimately leads to 10 possible 
different phases. However, as suggested by the snapshot configurations shown in Figure 3, only 4 phases are relevant. 

Figure 4 shows the probability of occurrence of the relevant phases as a function of the message intensity,  
corresponding to the small-$r$ regime (with $r=10^{-5}$) and for different 
values of the community connectivity: (a) $\nu=6\times 10^{-4}$ and (b) $\nu=2\times 10^{-3}$. In agreement 
with our qualitative discussion, Figure 4(a) shows that different kinds of asymmetric phase prevail when 
communities are loosely interconnected. Indeed, for small mass media fields, communities are predominantly 
in different non-$\mu$ states, i.e. the $(A_{non-\mu},B_{non-\mu})_A$ phase, 
while increasing $M$ above $M_c=7\times 10^{-4}$ the system is most often found in a 
bicultural phase with $\mu$ prevailing within community $A$, labeled as the $(A_\mu,B_{non-\mu})$ phase. 
Note also that, somewhat counterintuitively, 
the probability of achieving overall consensus tends to decrease as a function of the mass media intensity. 
This phenomenon is due to the fact that the external field prevents the independent auto-organization 
of the whole system in a non-$\mu$ state, while the lack of strong intercommunity ties prevent the message from 
reaching out far beyond the region directly exposed to the inhomogeneous mass media field.  
Figure 4(b) shows that, on the contrary, 
strongly interconnected communities allow the whole system to reach consensus. Increasing 
the mass media field, the system undergoes the expected transition from non-$\mu$ monocultural states, i.e. 
the symmetric $(A_{non-\mu},B_{non-\mu})_S$ phase, to $\mu$-consensus, indicated as $(A_\mu,B_\mu)$. 

\begin{figure}[t]
\begin{center}
\epsfxsize=3.8truein\epsfysize=2.7truein\epsffile{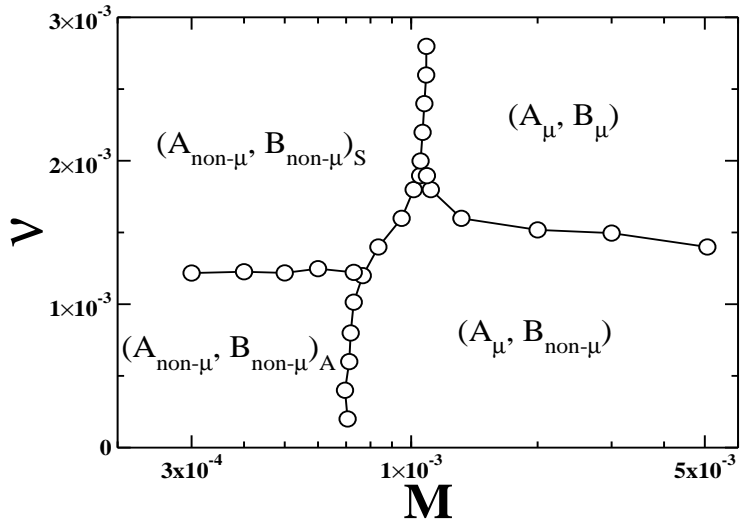}
\caption{$M-\nu$ phase diagram within the small-$r$ regime (for $r=10^{-5}$).}
\label{fig5}
\end{center}
\end{figure}

The corresponding phase diagram in the $M-\nu$ parameter space is shown in Figure 5. 
In the low-$M$ end, we observe that, increasing the intercommunity connectedness, the dominating phase changes 
from asymmetric non-$\mu$ to symmetric non-$\mu$. In fact, this transition matches the bicultural to monocultural 
phase transition observed earlier in the small-$r$ end in the absence of mass media fields (recall Figure 2). 
As discussed above, successful message spreading across the whole system can only be achieved when sufficiently 
strong mass media fields are applied on a sufficiently interconnected system. The boundaries for the $\mu$-consensus 
region are approximately $M\geq 10^{-3}$ and $\nu\geq 1.5\times 10^{-3}$. These results stress the fact that, in a 
general scenario, well-designed, cost-effective advertising campaigns should take into account the specific modular 
structure of the target population. Indeed, even very intense (and, hence, costly) mass media campaigns may fail if 
social modularity effects are disregarded. 

\begin{figure}[t]
\begin{center}
\epsfxsize=4.4truein\epsfysize=2.5truein\epsffile{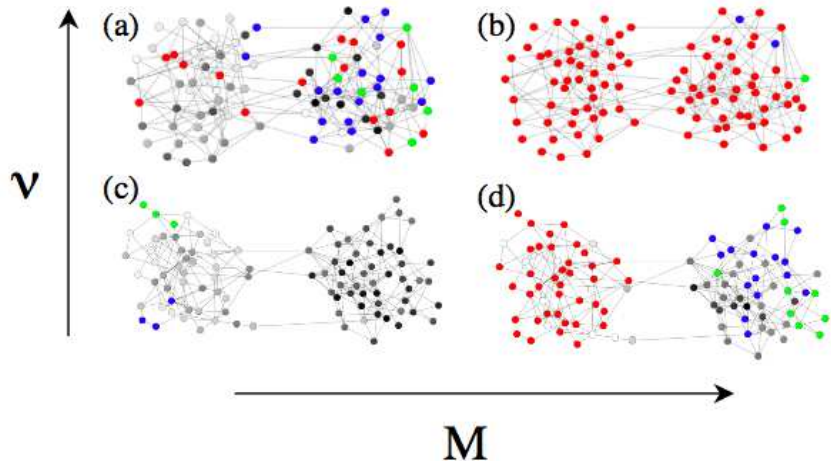}
\caption{Typical snapshot configurations for different values of $M$ and $\nu$ within the large-$r$ regime.
The $\mu$-state is shown in red, the most popular among non-${\bf\mu}$ states appears in blue, 
the second most popular non-${\bf\mu}$ state in green, while other states are shown in greyscale.}
\label{fig6}
\end{center}
\end{figure}

Considering networks of different size, phase diagrams 
can be made roughly invariant along the vertical axis by adopting the scaling relation $\nu N$. Indeed, this 
indicates that the relevant quantity defining the actual degree of interconnectedness is the total number 
of intercommunity links, $L_\nu$, in agreement with the theoretical argument presented above, Eq.(\ref{estimate}). 
The boundary between non-$\mu$ states and the $\mu-$consensus phase, which in Figure 5 appears around $M\simeq 10^{-3}$,    
is observed to shift towards lower values of $M$ as the network size is increased. We also explored the stability of our 
results under changes in the number of cultural attributes, $F$, and the number of different cultural traits per attribute, 
$q$, without noticing any significant variations. This behavior agrees well with previous investigations on 
Axelrod's model, where results roughly independent of the parameter $F$ \cite{cas00} and the parameter $q$ (for the 
model with noise, provided that $q\gg 1$) \cite{kle03b,maz07} were reported. A similar behavior is also observed in the 
large-$r$ regime, which is discussed below.
\begin{figure}[t]
\begin{center}
\epsfxsize=4.8truein\epsfysize=2.5truein\epsffile{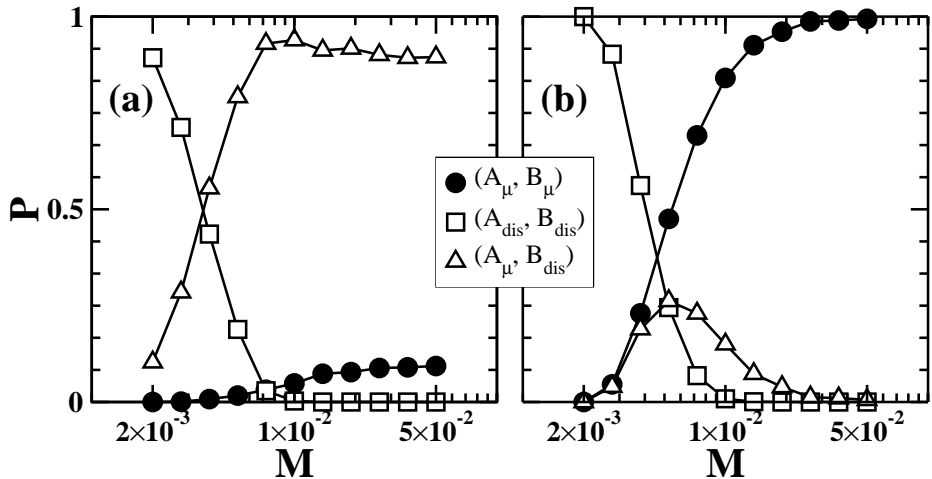}
\caption{Probability of occurrence of the relevant phases as a function of $M$ for  
the large-$r$ regime (with $r=10^{-3}$) and different 
values of the community connectivity: (a) $\nu=2\times 10^{-3}$ and (b) $\nu=1.7\times 10^{-2}$.}
\label{fig7}
\end{center}
\end{figure}
 
As anticipated, we now consider the large-$r$ scenario, which corresponds to a society characterized by loose social ties.      
Figure 6 shows typical snapshot configurations for different values of $M$ and $\nu$ within the large-$r$ regime ($r=10^{-2}$), 
with the same coloring scheme used in the visualizations of Figure 3. The corresponding parameter values are:  
(a) $M=1.5\times 10^{-2}, \nu=6\times 10^{-2}$; (b) $M=10^{-1}, \nu=6\times 10^{-2}$; (c) $M=1.5\times 10^{-2}, \nu=2\times 10^{-2}$; 
and (d) $M=10^{-1}, \nu=2\times 10^{-2}$.
 
Irrespective of connectivity, the low-$M$ region is characterized by disordered multicultural configurations 
(Figures 6(a) and 6(c)). Indeed, disordered states are characteristic of the large-$r$ region in the absence of mass 
media fields (compare to Figure 1(b)). Within this scenario,   
order can be achieved only when strong external fields oppose the large intrinsic noise (see Figure 6(b)). However, 
if the communities are sparsely interconnected, even strong fields are not capable of driving the system towards 
consensus: while the mass 
media message prevails within community $A$, the community $B$ is instead in a disordered multicultural state (Figure 6(d)).  

\begin{figure}[t]
\begin{center}
\epsfxsize=3.8truein\epsfysize=2.7truein\epsffile{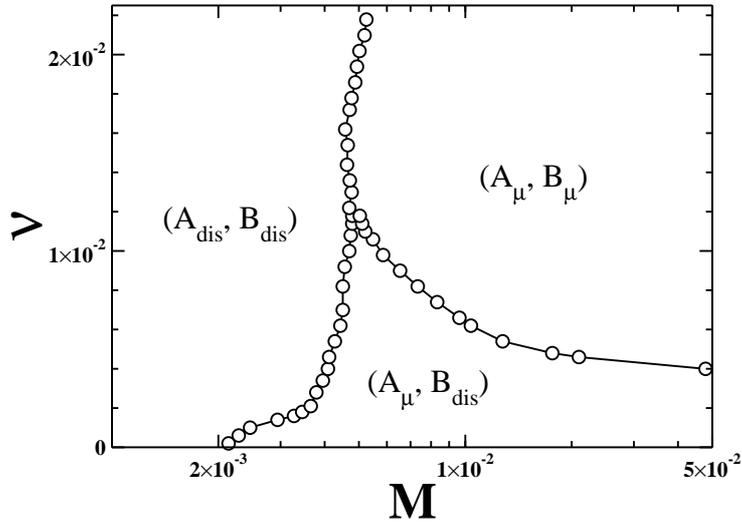}
\caption{$M-\nu$ phase diagram within the large-$r$ regime (for $r=10^{-3}$).}
\label{fig8}
\end{center}
\end{figure}

Following the procedure described above, we can determine the probability of occurrence of each phase as a function 
of the field intensity. However, out of 10 possible phases, only 3 are relevant.
These are shown in Figure 7 for $r=10^{-3}$ and different
values of the community connectivity: (a) $\nu=2\times 10^{-3}$ and (b) $\nu=1.7\times 10^{-2}$. 
As discussed above, when the communities are loosely interconnected (Figure 7(a)) the system is intrinsically 
disordered. Only a strong field can oppose the large noise level and order the community $A$, thus leading to 
a transition from the $(A_{dis},B_{dis})$ phase to the $(A_\mu,B_{dis})$ phase. For highly connected 
communities (Figure 7(b)), 
instead, the strong field is able to order the whole system in the $\mu$-state, driving the phase transition from 
$(A_{dis},B_{dis})$ to $(A_\mu,B_\mu)$.   

Figure 8 shows the phase diagram in the $M-\nu$ parameter space corresponding to the large-$r$ regime ($r=10^{-3}$).
Matching the large-$r$ region for the $M=0$ phase diagram of Figure 2, the low-$M$ end is dominated by 
the multicultural disordered phase. Increasing $M$, two different phases can be reached depending on the intercommunity 
connectedness: the $(A_\mu,B_{dis})$ phase, for loosely connected communities, and the $\mu$-consensus, when the 
communities are more strongly bound together. 

\begin{figure}[t]
\begin{center}
\epsfxsize=3.8truein\epsfysize=2.7truein\epsffile{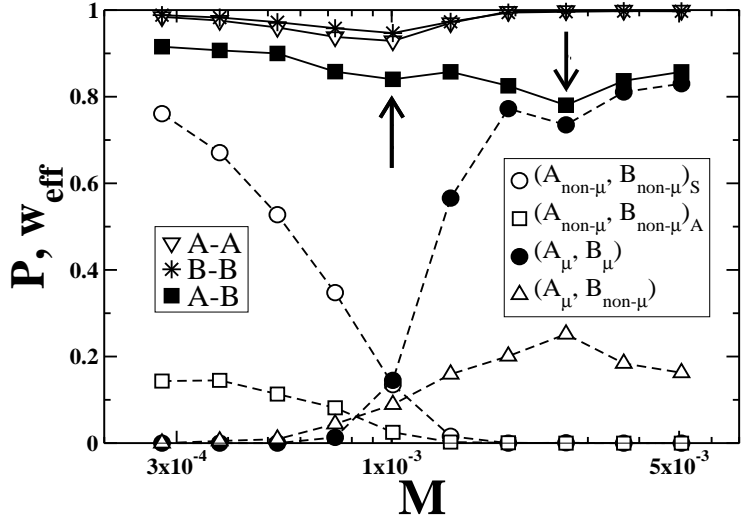}
\caption{Effective link weights as a function of $M$ 
for $r=10^{-5}$ and $\nu=2\times 10^{-3}$ (solid lines). Link weight plots are shown separately for intracommunity 
connections within each community, as well as for intercommunity connections. For comparison, 
the probability of occurrence of the relevant phases are also shown (dashed lines). See more details in the text.}  
\label{fig9}
\end{center}
\end{figure}

Finally, let us discuss some subtle, intriguing effects that result from the model dynamics. 
Recalling the phase probability distributions 
of Figure 4(b), the plot corresponding to the $(A_\mu,B_\mu)$ phase shows 
a dip at large message intensities, 
which is correlated with the occurrence of a bump in the plot of the $(A_\mu,B_{non-\mu})$ phase. Far from being an 
artifactual feature due to poor statistics, this phenomenon stems from the dynamical rules of the model and can be 
understood on the basis of a sound sociological interpretation. 
According to the second dynamical rule, within the community $A$, 
the parameter $M$ regulates the competition between 
two different types of interaction: that of an individual with the mass media, and that between neighboring individuals. 
This implies that, besides the ordering effect driven by the mass media interaction, 
which tends to align all cultural traits with the external message, 
there is also a competing disordering mechanism: individuals subject to strong mass media fields have a low 
probability of interacting (and, thus, of increasing the similarity) with their social neighbors. 
Although the former tends to prevail and is, ultimately, the mechanism  
responsible for the $\mu$-consensus ordering observed in the large-$M$ end of the phase diagrams, 
competition effects lead to visible features such as the dip and the bump noted above. 

In order to confirm this explanation, we computed the effective link weight, $w_{eff}(A-B)$, as the mean cultural 
overlap between neighbors that belong to different communities, and compared it to the effective link 
weight within each community. Figure 9 shows the effective link weights as a function of $M$ for the 
small-$r$ regime (with $r=10^{-5}$) and a large connectivity ($\nu=2\times 10^{-3}$). 
For the sake of comparison, the probability of occurrence of the relevant phases (same as those in Figure 4(b)) are also shown.   
Marked with arrows, we indicate two distinct features in the plots of effective link weight: a first dip (local minimum) 
in all three plots taking place at $M=10^{-3}$, and a second dip observed at $M=2.6\times 10^{-3}$ only in the plot 
of intercommunity links. 
The former is well correlated with the phase transition from 
$(A_{non-\mu},B_{non-\mu})_S$ to $(A_\mu,B_\mu)$, and hence reflects the corresponding phase changes within each community. 
Instead, the latter is well correlated with the dip in the probability of the $(A_\mu,B_\mu)$ phase, as well as with the 
bump in the probability of the $(A_\mu,B_{non-\mu})$ phase.

These results show that individuals subject to intense mass media fields are less 
likely to interact with their social neighbors, hampering message spreading processes  
across community boundaries. Thus,  
induced by mass media pervasiveness, the tendency of individuals towards isolated behavior is 
captured and well accounted for by this model.   
      
\section{Conclusions}

In the context of an extension of Axelrod's model for social influence, we studied  
how the modular structure of social 
networks affects the propagation of mass media messages across the society.  
The community structure of social networks was represented by coupled random networks, 
in which two random graphs are connected by intercommunity links. 

In the absence of mass media, 
we observed the prevalence of three distinct phases, depending on the values of cultural drift (i.e. the level 
of intrinsic noise) and community interconnectedness: 
the ordered monocultural phase, the ordered bicultural phase, and the disordered multicultural phase. 
The obtained phase diagram is qualitatively similar to that reported for the majority rule model defined 
on modular substrates \cite{lam07,lam07b}.   

Then, considering inhomogeneous advertising campaigns, we studied the 
system's response to the spreading of the mass media message. We considered separately two different scenarios: 
the small noise regime, which represents a society with individuals subject to 
strong social pressure, and the large noise regime, which is characterized by loose 
social ties.  
Incorporating the intensity of the mass media field as an additional parameter, we observed the emergence of 
new phases, which led to a very rich, novel phase diagram in the multidimensional parameter space.
Our results show that the design of successful, cost-effective advertising campaigns 
should take into account the specific modular 
structure of the target population. Indeed, even very intense (and, hence, costly) mass media campaigns may fail if 
social modularity effects are disregarded. 

Certainly, the simplified scenario considered here leaves room for further investigations that may look for 
modularity effects when communities of different size and different inter-/intra-community connectedness are considered.  
In this vein, the study of social influence and message spreading on real modular substrates taken from social interaction data 
(such as e.g. large-scale mobile phone usage with detailed space-time resolution \cite{can08,lam08}) could lead to interesting 
new results.   
We thus hope that the present findings will contribute to the growing interdisciplinary efforts 
in the mathematical modeling of social dynamics phenomena, and stimulate further work.

\section*{Acknowledgments}
We acknowledge the hospitality of the University of New Mexico, where this work was started during 
the authors' visit to the Consortium of the Americas for Interdisciplinary Science.  
J. C. is supported by the James S. McDonnell Foundation and the National 
Science Foundation ITR DMR-0426737 and CNS-0540348 within the DDDAS program.


\begin{thebibliography}{99}
\bibitem{oli99} de Oliveira S M, de Oliveira P M C and Stauffer D, 1999 
{\it Non-Traditional Applications of Computational Statistical Physics} (B.G. Teubner, Stuttgart)
\bibitem{wei00} Weidlich W, 2000 {\it Sociodynamics: a Systematic Approach to Mathematical Modelling in the 
Social Sciences} (Harwood Academic Publishers, Amsterdam) 
\bibitem{sta06} Stauffer D, de Oliveira S M, de Oliveira P M C and Sa Martins J S, 2006  
{\it Biology, Sociology, Geology by Computational Physicists} (Elsevier, Amsterdam)
\bibitem{gal00} Galam S and Wonczak S, 2000 Eur. Phys. J. B {\bf 18} 183
\bibitem{szn00} Sznajd-Weron K and Sznajd J, 2000 Int. J. Mod. Phys. C {\bf 11} 1157 
\bibitem{ber01} Bernardes A T, Costa U M S, Araujo A D and Stauffer D, 2001 
Int. J. Mod. Phys. C {\bf 12} 159 
\bibitem{ber02} Bernardes A T, Stauffer D and Kert\'esz J, 2002 Eur. Phys. J. B {\bf 25} 123
\bibitem{kup02} Kuperman M and Zanette D H, 2002 Eur. Phys. J. B {\bf 26} 387
\bibitem{ale02} Aleksiejuk A, Ho{\l}yst J A and Stauffer D, 2002 Physica A {\bf 310} 260
\bibitem{sch03} Schulze C, 2003 Int. J. Mod. Phys. C {\bf 14} 95
\bibitem{ben05} Ben-Naim E and Redner S, 2005 J. Stat. Mech. L11002
\bibitem{smi06} Smith R D, 2006 J. Stat. Mech. P02006
\bibitem{des06} De Sanctis L and Galla T, 2006 J. Stat. Mech. P12004
\bibitem{can06} Candia J, 2006 Phys. Rev. E {\bf 74} 031101 
\bibitem{can07a} Candia J, 2007 Phys. Rev. E {\bf 75} 026110 
\bibitem{ang07} Angelini L, Marinazzo D, Pellicoro M and Stramaglia S, 2007 J. Stat. Mech. L08001
\bibitem{bor07a} Bordogna C M and Albano E V, 2007 J. Phys.: Condens. Matter {\bf 19} 065144
\bibitem{bor07b} Bordogna C M and Albano E V, 2007 Phys. Rev. E {\bf 76} 061125
\bibitem{can07b} Candia J, 2007 J. Stat. Mech. P09001
\bibitem{cas07} Castellano C, Fortunato S and Loreto V, 2007 arXiv:0710.3256
\bibitem{axe97a} Axelrod R, 1997 J. Conflict Res. {\bf 41} 203
\bibitem{axe97b} Axelrod R, 1997 {\it The Complexity of Cooperation} (Princeton University Press, Princeton)
\bibitem{cas00} Castellano C, Marsili M and Vespignani A, 2000 Phys. Rev. Lett. {\bf 85} 3536 
\bibitem{kle03a} Klemm K, Egu\'{\i}luz V M, Toral R and San Miguel M, 2003 Phys. Rev. E {\bf 67} 026120 
\bibitem{kle03b} Klemm K, Egu\'{\i}luz V M, Toral R and San Miguel M, 2003 Phys. Rev. E {\bf 67} 045101(R) 
\bibitem{gon05} Gonz\'alez-Avella J C, Cosenza M G and Tucci K, 2005 Phys. Rev. E {\bf 72} 065102(R) 
\bibitem{gon06} Gonz\'alez-Avella J C, Egu\'{\i}luz V M, Cosenza M G,  Klemm K, Herrera J L and San Miguel M, 
2006 Phys. Rev. E {\bf 73} 046119 
\bibitem{kup06} Kuperman M N, 2006 Phys. Rev. E {\bf 73} 046139 
\bibitem{maz07} Mazzitello K, Candia J and Dossetti V, 2007 Int. J. Mod. Phys. C {\bf 18} 1475
\bibitem{gon07} Gonz\'alez-Avella J C, Cosenza M G,  Klemm K, Egu\'{\i}luz V M and San Miguel M, 
2007 J. Art. Soc. Soc. Simul. {\bf 10} 9
\bibitem{wat99} Watts D J, 1999 {\it Small Worlds} (Princeton University Press, Princeton)
\bibitem{alb02} Albert R and Barab\'asi A-L, 2002 Rev. Mod. Phys. {\bf 74} 47
\bibitem{new02} Newman M E J, Watts D J and Strogatz S H, 2002 Proc. Natl. Acad. Sci. USA {\bf 99} 2566 
\bibitem{dor03} Dorogovtsev S N and Mendes J F F, 2003 {\it Evolution of Networks} 
(Oxford University Press, New York)
\bibitem{new06} Newman M, Barab\'asi A-L and Watts D J (Eds.), 2006 {\it The Structure and Dynamics of Networks} 
(Princeton University Press, Princeton and Oxford) 
\bibitem{gir02} Girvan M and Newman M E J, 2002 Proc. Natl. Acad. Sci. USA {\bf 99} 7821
\bibitem{por06} Porter M A, Mucha P J, Newman M E J and Friend A J, 2007 Physica A {\bf 386} 414
\bibitem{pal07} Palla G, Barab\'asi A-L and Vicsek T, 2007 Nature {\bf 446} 664   
\bibitem{lam07} Lambiotte R and Ausloos M, 2007 J. Stat. Mech. P08026 
\bibitem{sha03} Shannon P, Markiel A, Ozier O, Baliga N S, Wang J T, Ramage D, Amin N, 
Schwikowski B and Ideker T, 2003 Genome Research {\bf 13} 2498
\bibitem{lam07b} Lambiotte R, Ausloos M and Ho{\l}yst J A, 2007 Phys. Rev. E {\bf 75} 030101(R)
\bibitem{can08} Candia J, Gonz\'alez M C, Wang P, Schoenharl T, Madey G and Barab\'asi A-L, 
2008 J. Phys. A: Math. Theor. {\bf 41} 224015  
\bibitem{lam08} Lambiotte R, Blondel V D, de Kerchove C, Huens E, Prieur C, 
Smoreda Z and Van Dooren P, 2008 arXiv:0802.2178
\end{thebibliography}
\end{document}